\documentclass[12pt]{iopart}
\usepackage{graphicx}
\newcommand{\be}{\begin{equation}}

\newcommand{\ee}{\end{equation}}

\begin{document}

\title{{Asymptotic and exact series representations for the incomplete Gamma function}}

\author{Paolo Amore\footnote{paolo@ucol.mx}}
\address{Facultad de Ciencias, Universidad de Colima, \\
Bernal D\'{\i}az del Castillo 340, Colima, Colima, M\'exico}

\begin{abstract}
Using a variational approach, two new series representations for the incomplete Gamma function are derived:
the first is an asymptotic series, which contains and improves over the standard asymptotic expansion;
the second is a uniformly convergent series, completely analytical, which can be used to obtain
arbitrarily accurate estimates of $\Gamma(a,x)$ for any value of $a$ or $x$. Applications of these formulas
are discussed.
\end{abstract}


\maketitle


\section{Introduction}

This paper is dedicated to the derivation of new series representations for the incomplete gamma function, 
which is given by the integral
\begin{eqnarray}
\Gamma(a,x) = \int_x^\infty t^{a-1} \ e^{-t} \ dt \ .
\label{eq_1}
\end{eqnarray}

The nice paper by Gautschi~\cite{GA} summarizes many properties of the incomplete gamma function and 
has been used as a valuable reference while writing this paper.
Since the incomplete gamma function occurrs in a variety of physical and engeneering problems, 
it is very valuable to obtain analytical series representations. 
Widely used is, for example, the asymptotic series~\cite{AS}
\begin{eqnarray}
\Gamma(a,x) &\approx& x^{a-1} \ e^{-x} \ \left[ 1 + \frac{a-1}{x} + \frac{(a-1) (a-2)}{x^2} + \dots \right] \ ,
\label{eq_2}
\end{eqnarray}
which can be applied to estimate $\Gamma$ for large $x$.
Other series representations, which are known in the literature, are valid either for small or large values of
$x$ (see \cite{AS}). The main purpose of this paper is to derive new series for the incomplete gamma function, which
can be used for arbitrary values of $x$ and $a$; such series could be very useful in dealing with a
range of applications in physics and engeneering  where the incomplete gamma function emerges.

The paper is organized as follows: in section \ref{sec_2} we describe our method and derive an asymptotic 
series for $\Gamma(a,x)$, which improves the estimates obtained with eq.~(\ref{eq_1}); in section \ref{sec_3} 
we generalize the results of section \ref{sec_2}, obtaining  a uniformly convergent series representation for 
$\Gamma(a,x)$, which can then be used to estimate $\Gamma(a,x)$ with arbitrary accuracy. We stress that all the formulas 
obtained in this paper are fully analytical. 
In section (\ref{sec_4}) we consider two applications of the incomplete gamma function,
and discuss the quality of the approximations obtained. Finally in section (\ref{sec_5}) we draw our conclusions.

\section{The method: asymptotic series}
\label{sec_2}

Techniques such as the Delta Expansion and Variational Perturbation Theory have been applied 
by many authors to physical problems of different nature, ranging from classical and quantum mechanics to
field theory (see for example \cite{lde,BMPS89,Jones:2000au,AFC90,Klei04} and references therein). 
In most cases these methods have allowed to obtain very useful and precise approximate solutions, 
which are valid even in the non-perturbative regime, where perturbation theory breaks down. 
The common idea behind these methods is simple and appealing: to introduce in the problem an arbitrary
parameter, to reorganize the expressions by identifying a leading and a ¨perturbative¨ term, both of which dependent
upon the arbitrary parameter, and finally to apply perturbation theory. To a finite perturbative order the
dependence upon the parameter survives and its effect is minimized by applying the Principle of 
Minimal Sensitivity (PMS)~\cite{Ste81}, which allows to obtain an optimal value of the parameter.
Since the parameter determined in this way turns out to be a
function of the coupling constants in the problem, the method yields a truly non-perturbative result, i.e. a result 
which is not polynomial in the coupling constants.

More recently  the author has applied these ideas to obtain a new series representation for the 
Riemann zeta function~\cite{Am04a} . The series which was found converges much faster than the 
standard series  $\zeta(s) = \sum\limits_{n=1}^\infty \frac{1}{n^s}$ 
and provides an analytical formula which is valid for arbitrary $s>0$. This formula has also been 
extended more recently to the critical line~\cite{Am04b}.

We now come to describe our method: with the purpose of introducing an arbitrary parameter 
we write the term $t^{a-1}$, appearing in the integral (\ref{eq_1}), as
\begin{eqnarray}
t^{a-1} = \sum_{r=0}^\infty \sum_{u=0}^r \frac{\Gamma(a)}{\Gamma(a-r) u! (r-u)!} (-1)^{r-u} \ \lambda^{a-u-1} \ t^u \ .
\label{eq_6}
\end{eqnarray}
$\lambda$ is the arbitrary parameter. Such expansion is valid as long as 
$\left|\frac{t-\lambda}{\lambda}\right|<1$, i.e. $0 < t <  2 \lambda$ and follows directly from the application
of the Binomial Theorem. 

Relaxing the constraint of convergence, we use eq.~(\ref{eq_6}) inside eq.~(\ref{eq_1}) and obtain
\begin{eqnarray}
\Gamma(a,x) \approx \sum_{r=0}^R \sum_{u=0}^r \frac{\Gamma(a)}{\Gamma(a-r) u! (r-u)!} (-1)^{r-u} \ \lambda^{a-1-u} \
 \int_x^\infty t^{u} \ e^{-t} \ dt  \ .
\label{eq_7}
\end{eqnarray}

Notice that we do not expect eq.~(\ref{eq_7}) to be an exact series representation of the incomplete gamma function,
since we have used eq.~(\ref{eq_6}) outside the region of convergence. For this reason we have used 
the symbol of approximation in eq.~(\ref{eq_7}) and we have substituted the upper limit in the sum over $r$, 
which should be infinite if the series were convergent, with a finite integer value $R$.

Although the integrals in  eq.~(\ref{eq_7}) are still incomplete gamma functions, 
the exponent $u$ is now integer and therefore such integrals can be calculated exactly
\begin{eqnarray}
\int_x^\infty t^{u} \ e^{-t} \ dt = e^{-x} \sum_{p=0}^u \frac{u!}{p!} x^p \ .
\label{eq_8}
\end{eqnarray}

Finally we can write eq.~(\ref{eq_7}) as
\begin{eqnarray}
\Gamma(a,x) \approx e^{-x} \ \sum_{r=0}^R \sum_{u=0}^r \sum_{p=0}^u \frac{\Gamma(a)}{\Gamma(a-r) p! (r-u)!} 
(-1)^{r-u} \ \lambda^{a-1-u} \ x^p  \ .
\label{eq_9}
\end{eqnarray}

Eq.~(\ref{eq_9}) constitutes a family of asymptotic series for the incomplete Gamma function:
if such series were convergent the arbitrary parameter $\lambda$ which appears explicitly in the equation, 
would disappear after performing the infinite sum. Although this is not true in the present case, 
given the asymptotic nature of the series, we decide to minimize the  dependence of the truncated series 
upon $\lambda$, as prescribed by the PMS. In \cite{Am04a}, for example, the PMS allowed to select a
series with a faster rate of convergence.

To first order we obtain the interesting result $\lambda_{PMS} = 1 + x$. We thus obtain the series 
\begin{eqnarray}
\Gamma(a,x) \approx e^{-x} \ \sum_{r=0}^R \sum_{u=0}^r \sum_{p=0}^u 
\frac{\Gamma(a)}{\Gamma(a-r) p! (r-u)!} (-1)^{r-u} \ (1+x)^{a-1-u} \ x^p  \ .
\label{eq_10a}
\end{eqnarray}

Remarkably, to first order, i.e. taking $R=1$,  eq.(\ref{eq_10a}) gives the approximate formula
\begin{eqnarray}
\Gamma(a,x) \approx e^{-x} \ (1+x)^{a-1} \ .
\label{eq_11}
\end{eqnarray}

\begin{figure}
\begin{center}
\includegraphics[width=8cm]{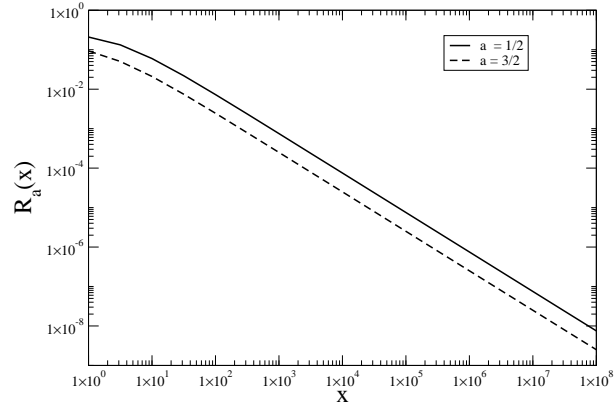}
\caption{The ratio $R_a(x)$ as a function of $x$ for fixed values of $a$.\label{fig_1_1}}
\end{center}
\end{figure}

In Fig.(\ref{fig_1_1}) our formula (\ref{eq_11}) has been compared with the leading order of eq.~(\ref{eq_2})
by considering the ratio
\begin{eqnarray}
R_a(x) \equiv \left|\frac{\Gamma_{1}(a,x)-\Gamma(a,x)}{\Gamma_{2}(a,x)-\Gamma(a,x)}\right| \ ,
\end{eqnarray}
where $\Gamma_1(a,x)$ is given by eq.~(\ref{eq_11}) and $\Gamma_1(a,x)$ is given by the leading order term
in eq.~(\ref{eq_2}). The better behaviour of our formula, eq.~(\ref{eq_11}) can be explained noticing that, 
by expanding it for large values of $x$, we reproduce correctly also the next to leading order in
eq.~(\ref{eq_2}).

\section{Convergent series}
\label{sec_3}

Although eq.~(\ref{eq_9}) provides a useful asymptotic representation for the incomplete gamma 
function, we are interested in obtaining an exact series representation for $\Gamma(a,x)$. 
Therefore we need to use the expansion eq.~(\ref{eq_6}) only within the region of convergence
and split the integral in eq.~(\ref{eq_1}) into two pieces:
\begin{eqnarray}
\Gamma(a,x) &=& \int_x^{\xi \lambda} t^{a-1} \ e^{-t} \ dt + \int_{\xi \lambda}^\infty t^{a-1} \ e^{-t} \ dt  \ .
\end{eqnarray}

The parameter $\xi$ in this expression, although otherwise arbitrary, must be chosen fullfilling the 
inequality $\frac{x}{\lambda} < \xi < 2$. In the first region one can apply the expansion (\ref{eq_6}) 
and obtain
\begin{eqnarray}
\Gamma(a,x) &=& \sum_{r=0}^\infty \sum_{u=0}^r \frac{\Gamma(a)}{\Gamma(a-r) u! (r-u)!} (-1)^{r-u} \ \lambda^{a-1-u} \
 \int_x^{\xi \lambda} t^{u} \ e^{-t} \ dt \nonumber \\
&+& \int_{\xi \lambda}^\infty t^{a-1} \ e^{-t} \ dt 
= \sum_{r=0}^\infty \sum_{u=0}^r \frac{\Gamma(a)}{\Gamma(a-r) u! (r-u)!} (-1)^{r-u} \ \lambda^{a-1-u} \nonumber \\
&\cdot& \left[ e^{-x} \sum_{p=0}^u \frac{u!}{p!} x^p - e^{-\xi \lambda} \sum_{p=0}^u \frac{u!}{p!} (\xi \lambda)^p
\right] + \Gamma(a,\xi \lambda)
\end{eqnarray}

As a result, the incomplete gamma function, evaluated at a given point $x$ is obtained in terms of the 
the incomplete gamma function, evaluated at a larger point $\xi \lambda$, which lies within a maximum distance of 
$2 \lambda-x$. This procedure can now be easily iterated by resorting to additional arbitrary parameters, but  
we prefer to fix $\lambda$ to the optimal value found previously, i.e. $\lambda = \lambda_{PMS} = 1+x$ and to define 
$x_0 \equiv x$, $x_1 \equiv  \xi \lambda = \xi (1+x_0)$, $\dots$, 
$x_q \equiv \xi (1+x_{q-1})$. 

We then obtain
\begin{eqnarray} 
\Gamma(a,x) &=& \sum_{q=0}^\infty \sum_{r=0}^\infty \sum_{u=0}^r \frac{\Gamma(a)}{\Gamma(a-r) u! (r-u)!} (-1)^{r-u} \ 
(1+x_q)^{a-1-u} \nonumber \\
&\cdot& \sum_{p=0}^u \frac{u!}{p!} \left[ e^{-x_q}  x_q^p - e^{-\xi (1+x_q)} (\xi (1+x_q))^p\right] \ ,
\end{eqnarray}
a uniformly convergent series representation for $\Gamma(a,x)$, which depends upon an arbitrary parameter $\xi$.
Notice that $x_q = \xi^q \ x + \xi \frac{\xi^q-1}{\xi-1}$. 
Although one could invoke once more the PMS and optimize the series with respect to $\xi$, a large gain in simplicity
comes from considering $\xi = 1$, for which one obtains $x_q = x +q$.

In this limit we obtain the series
\begin{eqnarray} 
\Gamma(a,x) &=& \sum_{q=0}^\infty \sum_{r=0}^\infty \sum_{u=0}^r  \sum_{p=0}^u \frac{1}{p!} 
\frac{\Gamma(a)}{\Gamma(a-r)  (r-u)!} (-1)^{r-u} \ (1+x + q)^{a-1-u} e^{-x-q} \nonumber \\
&\cdot& \left(  (x+q)^p - \frac{1}{e} \ (1+x+q)^p\right)
\label{eq_14}
\end{eqnarray}

Eq.~(\ref{eq_14}) constitutes the main result of this paper: when the sums over $r$ and $q$ are truncated to a finite
order this equation provides an analytical approximation to the exact value; the precision of the 
approximation will be arbitrarily high for sufficiently large cutoff values of $r$ and $q$. 
We will call the truncated sum $\Gamma_{(r_{max},q_{max})}(a,x)$ and define the ``reduced'' function
\begin{eqnarray}
\tilde{\Gamma}_{(r_{max},q_{max})}(a,x) = \frac{\Gamma_{(r_{max},q_{max})}(a,x)}{e^{-x} (1+x)^{a-1}}
\end{eqnarray}
where the leading asymptotic behaviour, given by eq.~(\ref{eq_11}) has been taken out.

\begin{figure}
\begin{center}
\includegraphics[width=14cm]{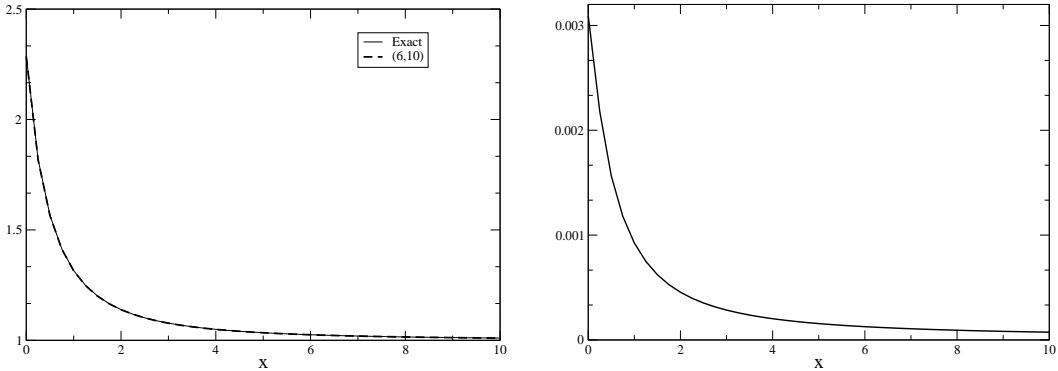}
\caption{Left: Comparison between $\tilde{\Gamma}(\pi,x)$ and  $\tilde{\Gamma}_{(6,10)}(\pi,x)$;
Right: $\tilde{\Gamma}(\pi,x)-\tilde{\Gamma}_{(6,10)}(\pi,x)$. 
\label{fig_1_2}}
\end{center}
\end{figure}

In the left plot of figure \ref{fig_1_2} we compare the numerical values obtained for 
$\tilde{\Gamma}(\pi,x)$ (solid thin line) with the results obtained using the analytical formula 
corresponding to  $\Gamma_{(6,10)}(\pi,x)$.
In the right plot we plot the difference between the two curves: the maximum error is found for $x=0$.

By taking the limit $x \rightarrow 0$ in eq.~(\ref{eq_14}) we also obtain the identity
\begin{eqnarray} 
1 &=& \sum_{q=0}^\infty \sum_{r=0}^\infty \sum_{u=0}^r  
\frac{1}{\Gamma(a-r) (r-u)!} (-1)^{r-u} \ (1+ q)^{a-1-u} e^{-q} \left(  1 - \frac{1}{e} \right) 
\nonumber \\
&+& \sum_{q=0}^\infty \sum_{r=0}^\infty \sum_{u=0}^r  \sum_{p=1}^u \frac{1}{p!} 
\frac{1}{\Gamma(a-r) (r-u)!} (-1)^{r-u} \ (1+ q)^{a-1-u} e^{-q} \left(  q^p - \frac{1}{e} \  
(1+q)^p\right) \nonumber
\label{eq_15}
\end{eqnarray}
which we have verified numerically\footnote{Some care should be used  because of 
the condition of convergence $0 < t < 2 \lambda$.}.

Notice that to improve the accuracy of the  series for small $x$, one could also split the integral 
and write
\begin{eqnarray}
\Gamma (a,x) &=& \int_x^\epsilon t^{a-1} \ e^{-t} \ dt + \int_\epsilon^\infty t^{a-1} \ e^{-t} \ dt .
\end{eqnarray}
The first integral could be evaluated with high precision Taylor expanding the exponential to a finite
order , provided that $x< \epsilon \approx 1$, whereas the second integral could be evaluated using 
once again our formula.

\section{Applications}
\label{sec_4}

A first application of the results obtained in this paper concerns the calculation of the probability integral 
\begin{eqnarray}
P(x) = \frac{1}{\sqrt{2 \pi}} \ \int_0^x e^{-t^2/2} dt = \frac{1}{2} \ {\rm erf\left(\frac{x}{\sqrt{2}}\right)}
\label{eq_16}
\end{eqnarray}
which can be related to the incomplete gamma function by
\begin{eqnarray}
{\rm erf} \ x = 1 - \frac{\Gamma\left(\frac{1}{2},x^2\right)}{\sqrt{\pi}} \ .
\end{eqnarray}

In \cite{Bag95} Bagby provides a simple approximation to $P(x)$, given by:
\begin{eqnarray}
P(x) \approx \frac{1}{2} \left\{ 1 - \frac{1}{30} \left[ 7 \ e^{-x^2/2} + 16 \ e^{-x^2 (2-\sqrt{2})} + 
\left(7+\frac{\pi x^2}{4} \right)\ e^{-x^2} \right]\right\}^{1/2} \ .
\end{eqnarray}

\begin{figure}
\begin{center}
\includegraphics[width=8cm]{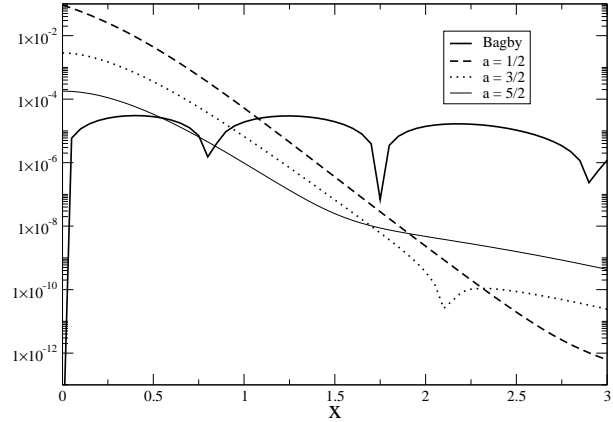}
\caption{The error over identity of eq.~(\ref{eq_16}) as a function of $a$, using the cutoffs in 
$r_{max} = 10$ and in $q_{max}=20$.\label{fig_1_4}}
\end{center}
\end{figure}

In Fig.~\ref{fig_1_4} we plot the error over $P(x)$, defined as $\left| P(x) - P_{approx}(x)\right|$, using the 
formula of Bagby and our formula eq.~(\ref{eq_14}) taking $r_{max} = 10$ and $q_{max} = 20$ and relating 
$P(x)$ to $\Gamma(1/2,x^2/2)$, $\Gamma(3/2,x^2/2)$ and $\Gamma(5/2,x^2/2)$ respectively.
Our analytical formula can be made as precise as we wish, both by raising the values of the cutoffs in the 
series and by considering higher orders in the incomplete gamma function. 

\begin{figure}
\begin{center}
\includegraphics[width=14cm]{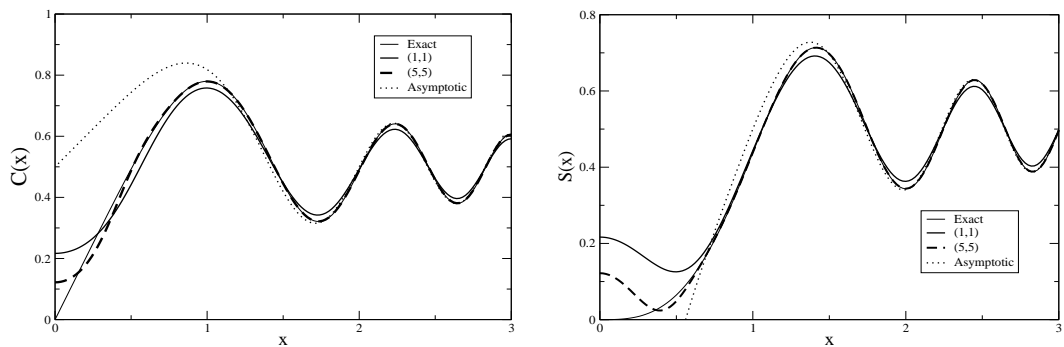}
\caption{Fresnel integrals $C(x)$ (left plot) and $S(x)$ (right plot) calculated exactly (solid line), using
the asymptotic formula of eq.~(\ref{asym}) and using eq.~(\ref{eq_14}) to order $(1,1)$ and $(5,5)$. \label{fig_1_5}}
\end{center}
\end{figure}

Another application of our formula is the calculation of Fresnel integrals~\cite{GA}, which naturally occurr 
in the treatment of diffraction problems:
\begin{eqnarray}
C(x) = \int_0^x \ \cos \left( \frac{\pi}{2} t^2\right) \ dt \ \ \  , \ \
S(x) = \int_0^x \ \sin \left( \frac{\pi}{2} t^2\right) \ dt \ .
\end{eqnarray} 
These integrals can be related to the error function by means of the equation
\begin{eqnarray}
C(x) + i \ S(x) = \frac{1+i}{2} \ {\rm erf}(z)
\end{eqnarray}
where $z = \frac{\sqrt{\pi}}{2} \ (1-i) \ x$. Our formula already to low order 
provides an excellent {\sl analytical} approximation to the Fresnel integrals, with somewhat larger 
errors for small $x$ (see Figure \ref{fig_1_5}).
Such errors decrease when higher order contributions are considered. The dotted lines correspond 
to the asymptotic formulas:
\begin{eqnarray}
C(x) \approx \frac{1}{2} + \frac{\sin \frac{\pi x^2}{2}}{\pi x} \ \  \ , \ \ \
S(x) \approx \frac{1}{2} - \frac{\cos \frac{\pi x^2}{2}}{\pi x}  \  .
\label{asym}
\end{eqnarray}

\section{Conclusions}
\label{sec_5}

In this paper we have considered the incomplete gamma function and obtained two new series representations:
an asymptotic series, which improves over the standard asymptotic expansion of $\Gamma(a,x)$, and a convergent
series, which to lowest orders can be used to obtain arbitrarily precise analytical approximations to the 
incomplete gamma function. As an application of the results obtained in this way, we have used our analytical formula 
to calculate the probability integral and we have compared the results with a particular approximation~\cite{Bag95}. 
By working to order $(1,1)$ and $(5,5)$ we also have found a very 
precise {\sl analytical} approximation to the Fresnel integrals, which occurr in the treatment of diffraction.
Such approximation could be further improved by considering the contributions of higher orders.

The generality of the ideas upon which the method that we have propesed relies and the previous success in 
dealing with the Riemann and Hurwitz zeta functions motivates the effort to obtain similar convergent 
series representation for other special function. Work in this direction is currently in progress.

\verb''\ack
 The author thanks Ricardo A. Saenz for useful conversations.
Support of Conacyt grant no. C01-40633/A-1 and of Fondo Ramon Alvarez-Buylla of the Universidad de Colima is 
also acknowledged.

\verb''\section*{References}
\verb''

\end{document}